# Negative reflection of polaritons at the nanoscale in a low-loss natural medium

Gonzalo Álvarez-Pérez[1,2,†], Jiahua Duan[1,2,†,*], Javier Taboada-Gutiérrez[1,2], Qingdong Ou[3], Elizaveta Nikulina[4], Song Liu[5], James H. Edgar[5], Qiaoliang Bao[3], Vincenzo Giannini[6,7,8], Rainer Hillenbrand[4,9], J. Martín-Sánchez[1,2], Alexey Y. Nikitin[9,10,*], Pablo Alonso-González[1,2,*]

[1]Department of Physics, University of Oviedo, Oviedo 33006, Spain.
[2]Center of Research on Nanomaterials and Nanotechnology, CINN (CSIC-Universidad de Oviedo), El Entrego 33940, Spain.
[3]Department of Materials Science and Engineering, and ARC Centre of Excellence in Future Low-Energy Electronics Technologies (FLEET), Monash University, Clayton, Victoria, Australia.
[4]CIC nanoGUNE BRTA and Department of Electricity and Electronics, UPV/EHU, Donostia/San Sebastián 20018, Spain.
[5]Tim Taylor Department of Chemical Engineering, Kansas State University, Manhattan, KS 66506, USA.
[6]Instituto de Estructura de la Materia (IEM), Consejo Superior de Investigaciones Científicas (CSIC), Serrano 121, Madrid 28006, Spain
[7]Technology Innovation Institute, Building B04C, Abu Dhabi P.O. Box 9639, United Arab Emirates
[8]Centre of Excellence ENSEMBLE3 sp. z o.o., Wolczynska 133, Warsaw, 01-919, Poland.
[9]IKERBASQUE, Basque Foundation for Science, Bilbao 48013, Spain.
[10]Donostia International Physics Center (DIPC), Donostia/San Sebastián 20018, Spain
*pabloalonso@uniovi.es, alexey@dipc.es, duanjiahua@uniovi.es

[†]These authors contributed equally to this work.

**Negative reflection occurs when light is reflected towards the same side of the normal to the boundary from which it is incident. This exotic optical phenomenon, which provides a new avenue towards light manipulation, is not only yet to be visualized in real space but remains largely unexplored both at the nanoscale and in natural media. Here, we directly visualize nanoscale-confined polaritons negatively reflecting on subwavelength mirrors fabricated in a low-loss van der Waals crystal. Our near-field nanoimaging results unveil an unconventional and broad tunability of both the polaritonic wavelength and direction of propagation upon negative reflection. Based on these findings, we introduce a novel device in nano-optics: a hyperbolic nanoresonator, in which hyperbolic polaritons with different momenta reflect back to a common point source, enhancing its intensity. These results pave the way to realize nanophotonics in low-loss natural media, providing a novel and efficient route to confine and control the flow of light at the nanoscale, key for future optical on-chip nanotechnologies.**

The control and manipulation of light and light-matter interactions at the nanoscale, the main aim of the field of nano-optics[1], is central to the development of next-generation nano-optical devices. In this regard, extensive efforts have been directed towards the understanding of fundamental optical phenomena at the nanoscale, such as reflection and refraction, even more so in their most exotic and counterintuitive versions: anomalous and negative reflection and refraction. For instance, planar lenses have been developed based on anomalous refraction; and nano-light has been proved to reflect anomalously in nanocones made of boron nitride[2] and to be edge-steerable along counterintuitive directions in microslits made of molybdenum trioxide[3], enabling novel avenues in nano-photonics. By contrast, negative reflection, which may provide another path towards manipulating light-matter excitations at the nanoscale, remains explored to a lesser extent. To date, this exotic phenomenon has only been observed for free-space light reflecting off artificially engineered interfaces (metasurfaces)[4-8] and chiral mirrors[9]. A key disadvantage of these platforms is that they preclude the direct, real-space visualization of the phenomenon, which has been exceptionally difficult to accomplish experimentally, as not only low optical losses are required to guarantee a clear real-space visualization, but also an adequate material platform. This real-space visualization can provide crucial insights for a comprehensive understanding of such a fundamental optical phenomenon as reflection, opening so far unexplored possibilities in nano-optics.

The recent findings of HPhPs in low-dimensional van der Waals (vdW) crystals[10-18] have provided low-loss natural media to explore exotic optical phenomena at the nanoscale[19-29]. In particular, HPhPs in α-MoO$_3$ are ideal candidates for such optical studies since they guarantee a direct experimental visualization along the surface due to their in-plane hyperbolic dispersion, as recently demonstrated in planar refraction and focusing studies[27-29]. Here, we demonstrate real-space visualization of negative reflection of polaritons at the nanoscale and in a low-loss vdW crystal (α-MoO$_3$), unveiling a non-intuitive, unconventional and broad tunability of the polaritonic wavelength and direction of propagation, providing fundamental insights on reflection of nanolight. To do this, we introduce an optical scheme that combines near-field interferometry measurements of HPhPs in α-MoO$_3$ (employing scattering-type near-field optical microscopy, s-SNOM, see Methods) with the design and fabrication of tilted edges with subwavelength dimensions acting as mirrors.

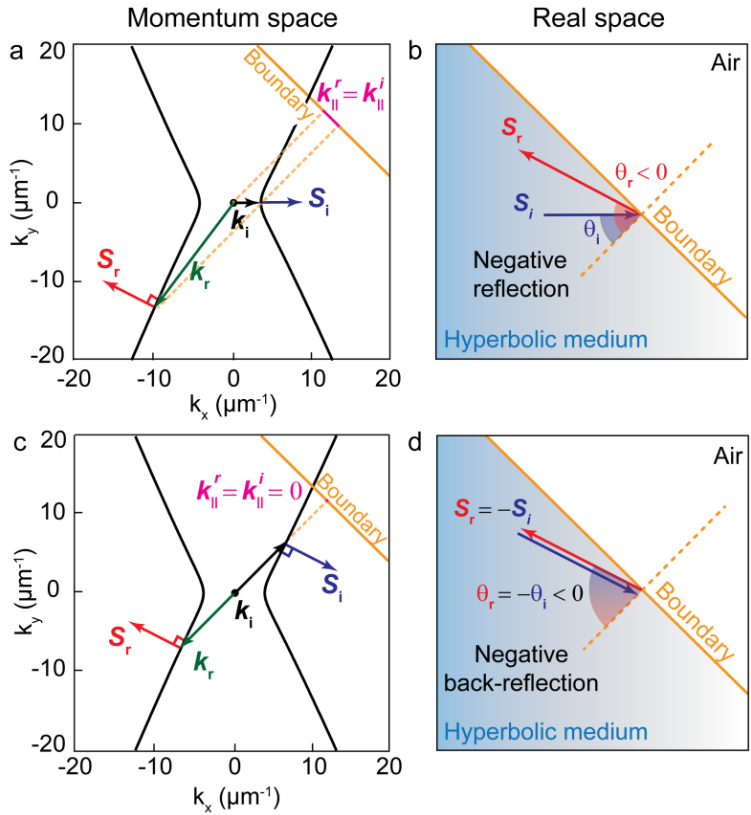

**Fig. 1 | Schematics of negative reflection of HPhPs.** (**a**) Negative reflection of HPhPs upon momentum conservation at the boundary ($\Delta k_\parallel = k_\parallel^r - k_\parallel^i = 0$). The black solid line represents the IFC of HPhPs. $\boldsymbol{k}_{i,r}$ and $\boldsymbol{S}_{i,r}$, generally non-collinear, indicate the wavefronts and Poynting vectors of the incident/reflected HPhPs, respectively. (**b**) Real-space illustration of negative reflection of HPhPs. The angle of reflection $\theta_r$ is negative and $\boldsymbol{S}_i$ and $\boldsymbol{S}_r$ are on the same side of the normal to the boundary. (**c**) Negative back-reflection of HPhPs ($k_\parallel^r = k_\parallel^i = 0$), yielding $\boldsymbol{k}_i = -\boldsymbol{k}_r$ non-collinear with $\boldsymbol{S}_i = -\boldsymbol{S}_r$. (**d**) Real space illustration of negative back-reflection of HPhPs, yielding only a single beam off the normal.

Our finding relies on the exotic properties of HPhPs, which stem from the anisotropic crystal structure of the host medium, giving rise to different optical responses along different crystal directions[30]. The anisotropy of HPhPs is captured by their hyperbolic iso-frequency curve (IFC, Fig. 1a), a slice of the dispersion surface in wavevector-frequency ($\boldsymbol{k}\omega$) space —where $\boldsymbol{k} = (k_x, k_y)$ is the polaritonic in-plane wavevector— by a plane of a constant frequency $\omega$. The Poynting vector $\boldsymbol{S}$, which determines the direction of the polaritonic energy flux, and thus its propagation direction, is normal to the IFC. As such, $\boldsymbol{k}$ and $\boldsymbol{S}$ are generally non-collinear in hyperbolic media, which has dramatic consequences in reflection phenomena. Specifically, upon momentum conservation at the boundary (the projection of the incident and reflected wavevectors on the boundary, $k_\parallel^{i,r}$, must be conserved, i.e., $\Delta k_\parallel = k_\parallel^r - k_\parallel^i = 0$), HPhPs can reflect at angles which are different to that of incidence, and the angle of reflection ($\theta_r$, calculated as the angle that the reflected energy flux $\boldsymbol{S}$ forms with the normal to the boundary) can be even negative, i.e., $\boldsymbol{S}_i$ and $\boldsymbol{S}_r$ can be on the same side of the normal to the boundary (Fig. 1b). This is in stark contrast with the behaviour in isotropic media, where the angle of reflection is positive and equal to that of incidence $\theta_r = \theta_i$ (specular reflection). In addition, the wavevector of the outcoming wave, $k_r$, can be much greater in modulus than that of incidence $k_i$, implying that the reflected wave can exhibit a much smaller wavelength, thereby providing a means to squeeze nano-light well beyond the diffraction limit by reflection. Remarkably, the phenomenon of negative reflection can be clearer pictured under the condition of back-reflection. When the incident and reflected wavevectors are both normal to the reflecting boundary, i.e., $k_\parallel^r = k_\parallel^i = 0$ (Fig. 1c), then the energy is reflected back to the source, i.e., $\boldsymbol{S}_r = -\boldsymbol{S}_i$ and $\theta_r = -\theta_i < 0$ (Fig. 1d). As such, the incident and reflected beams propagate along the same direction, thereby interfering and giving rise to a neat picture of negative reflection, with only a single interference beam off the normal.

Crucially, back-reflection is the optical scheme used in s-SNOM polariton interferometry[31,32] (Methods), where near-field images are obtained by raster scanning a metallic tip that acts both as source and detector of polaritonic fields (Fig. 2a). Therefore, to experimentally demonstrate negative reflection at the nanoscale and in a natural medium, we performed s-SNOM polariton interferometry in α-MoO₃, which supports nanoscale-confined, low-loss in-plane HPhPs in the mid-infrared range[33]. Specifically, we study HPhPs back-reflecting at edges fabricated in thin slabs of α-MoO₃ by focused ion beam (Methods). By carefully selecting the angle $\varphi$ that the edges form with respect to the [100] crystal axis of α-MoO₃, and, importantly, by fabricating them with subwavelength dimensions, we can unambiguously probe negative back-reflection of HPhPs along specific directions, avoiding the interference with HPhPs propagating along other directions (the use of large edges acting as mirrors would hinder the identification of the reflection direction due to interference effects, see Fig. S4 in Supplementary Information). In addition, the subwavelength edges are slightly rounded at their terminations to avoid the strong launching of HPhPs, which would also produce interferences that would distort our real-space visualization of negative reflection.

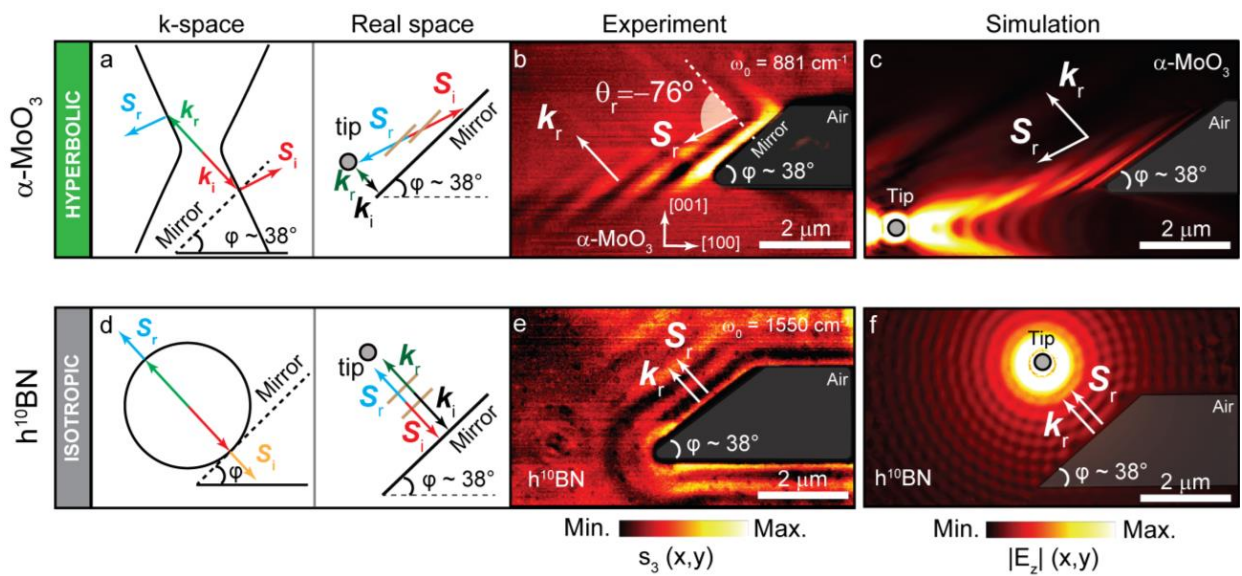

**Fig. 2 | Visualization of negative reflection of nanoscale-confined polaritons in a natural medium.** Illustration of back-reflection of HPhPs in k-space and real space for (**a**) in-plane hyperbolic and (**d**) in-plane isotropic media. (**b,e**) Experimental s-SNOM amplitude image $s_3(x,y)$ of HPhPs back-reflecting on mirrors tilted at an angle $\varphi = 38°$ fabricated on α-MoO$_3$ and h$^{10}$BN, respectively. $k_{i,r}$ and $S_{i,r}$ indicate the wavefronts and Poynting vectors of the incident/reflected HPhPs, respectively. (**c,f**) Full-wave numerical simulation $E_z(x,y)$ of back-reflected polaritons excited by a point dipole on α-MoO$_3$ and h$^{10}$BN, respectively.

As a representative example, we fabricate edges tilted at an angle $\varphi = 38°$, thereby probing polaritonic states close to asymptote of the hyperbolic IFC, where the non-collinearity between $k$ and $S$ is maximal (Fig. 2a). By s-SNOM imaging, we observe interferometric fringes revealing that HPhPs anomalously back-reflect at counterintuitive directions (Fig. 2b). In particular, the Poynting vector of the back-reflected wave (indicated by the direction of the maxima of the near-field amplitude[3,10]) is along the same side of the normal to the boundary as the incident one, unambiguously demonstrating negative reflection of HPhPs at an angle $\theta_r = -76°$. In addition, the back-reflected wavefronts appear tilted with respect to the direction of propagation and almost parallel to it (see Supplementary Sections S1,2), which is explained by $k$ and $S$ being almost perpendicular at some points of the hyperbolic IFC (Fig. 2a). To corroborate our experimental images and their origin in the negative reflection of PhPs, we perform full-wave simulations mimicking the experiment (see Methods). Fig. 2c represents the distribution of the electric field, |Ez|, created by a vertical point dipole (representing the experimental tip, see Methods) at a fixed lateral position (chosen along the expected direction of negatively reflected PhPs according to Fig. 2a). We can clearly recognize similar tilted interference fringes as those observed in the near-field image, thus unambiguously demonstrating their origin in the back-reflection of PhPs at the mirror. For comparison, we also visualize in-plane isotropic polaritons back-reflecting on mirrors[17] fabricated in isotopically enriched h$^{10}$BN (Fig. 2d). As expected, we observe interference fringes parallel to all the mirrors (Fig. 2e) and thus along all in-plane directions, in excellent agreement with our simulations (Fig. 2f).

To gain further insights into the negative reflection at the nanoscale, we study the dependence of the polaritonic wavelength $\lambda_p$ and the angle of reflection $\theta_r$ as a function of the angle of the mirror $\varphi$ and incident frequency $\omega_0$. To do so, we fabricate mirrors in α-MoO$_3$ at different angles $\varphi = 38º$, 60º and 90º and perform s-SNOM polariton interferometry of back-reflected HPhPs. Since in anisotropic media the wavevector is direction-dependent, $\lambda_p$ changes depending on the direction along which we map back-reflection, and hence on $\varphi$. By selecting $\varphi$, we probe the behaviour of HPhPs upon negative reflection for different **k** and **S**, i.e., different angles of incidence. This is illustrated in Figs. 3A, B for $\omega_0 = 889$ cm$^{-1}$.

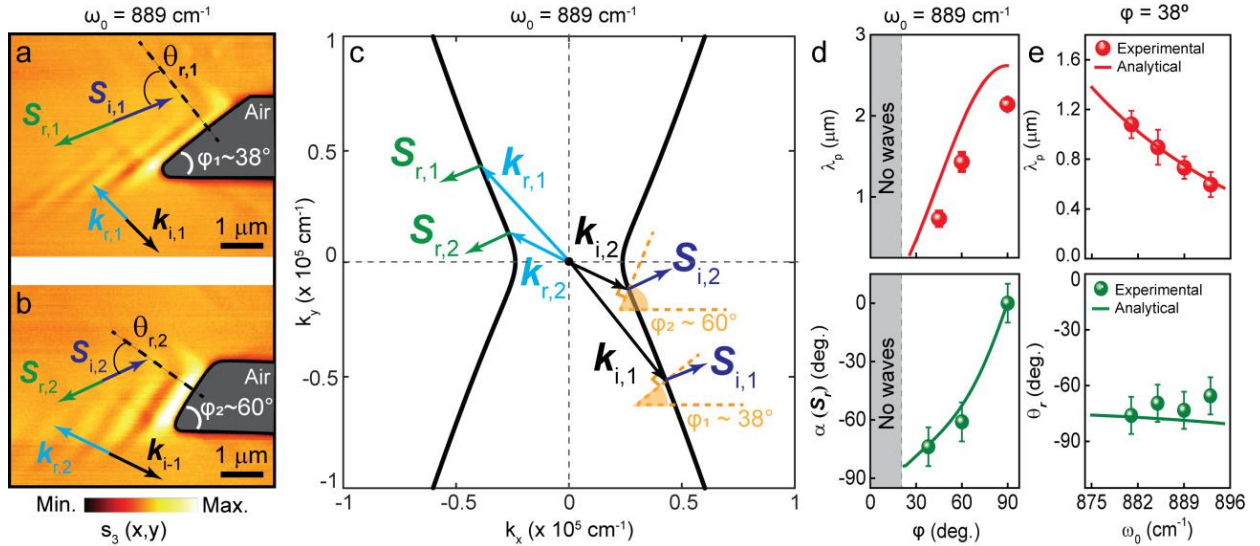

**Fig. 3 | Properties of negative reflection of HPhPs in α-MoO$_3$.** (**a, b**) Experimental near-field images $s_3(x,y)$ of HPhPs back-reflecting on mirrors fabricated in α-MoO$_3$ tilted at $\varphi = 38º$ and 60º, respectively. (**c**) IFC of HPhPs showing the directions of $k_{i,r}$ and $S_{i,r}$ of the incident and back-reflected HPhPs on mirrors tilted at $\varphi = 38º$ and 60º. (**d, e**) Dependence of the polaritonic wavelength $\lambda_p$ (upper panel) and angle of reflection $\theta_r$ (lower panel) on $\varphi$ ($\omega_0 = 889$ cm$^{-1}$) and $\omega_0$ ($\varphi = 38º$), respectively. The continuous lines in panels D and E are extracted from analytical calculations and the symbols from experimental s-SNOM images.

Clearly, $\lambda_p$ and $\theta_r$ change as a function of $\varphi$. Specifically, as a result of the hyperbolic-like shape of the IFC, when $\varphi$ decreases (from 90º to 20º), $k_i$ and $k_r$ approach the asymptote (see black and cyan arrows in Fig. 3a,b,c), and $\lambda_p$ can be dramatically squeezed (from 2.5 to 0 μm, Fig. 3d, top panel), while, on the other hand, $\theta_r$ takes values in a very broad angular sector (from 0º to -80º, Fig. 3d, bottom panel). Note that $\varphi$ values smaller than 20º are forbidden since the corresponding wavevectors are not allowed by the hyperbolic IFC (grey shaded area). The continuous lines show calculations based on the analytical IFC and momentum conservation at the mirror, in excellent agreement with the experiment. Taken together, these results show that both $\lambda_p$ and $\theta_r$ exhibit an unconventional and broad tunability as a function of $\varphi$, in stark contrast to back-reflected polaritons in isotropic media, where the same $\lambda_p$ and $\theta_r$ is obtained for all $\varphi$. We note that polariton interferometry of back-reflected polaritons is the technique that allows us to probe these exotic aspects of negative reflection, yet the conclusions are broader and apply beyond the case of back-reflection and even of negative reflection, providing a novel strategy to squeeze and steer nano-light.

Given the strong dispersive nature of the back-reflected polaritons, we further study their dependence upon $\omega_0$. The upper and lower panels in Fig. 3e represent $\lambda_p$ and $\theta_r$ as a function of $\omega_0$, respectively, together with analytical calculations (green continuous lines), showing an excellent agreement. The spectral dependence demonstrates that $\lambda_p$ and $\theta_r$ can be tuned between 0.6 and 1.1 µm and -76° and -65°, respectively, when $\omega_0$ varies from 881 to 893 cm$^{-1}$. Note that this dependence does not hold in isotropic media, where $\theta_r$ is always equal to zero under back-reflection regardless of $\omega_0$.

Following our findings and basic understanding on negative back-reflection, we further derive the analytical shape of a polaritonic mirror capable of reflecting all the incident HPhPs back to their source, i.e., a hyperbolic retroreflector. Such curve must meet the condition imposing that every incident wavevector is perpendicular to it. In isotropic media this condition is satisfied by a circumference (inset in Fig. 4a). However, in hyperbolic media, due to the non-collinearity of $\boldsymbol{k}$ and $\boldsymbol{S}$, the condition yields the following hyperbola (Supplementary Section S6):

$$y(x) = \pm \sqrt{\frac{x^2 |\varepsilon_y|}{|\varepsilon_x|} - a^2}, \qquad (1)$$

where $x$ and $y$ are the directions along the main and conjugate axes of the hyperbola, $\varepsilon_x$ and $\varepsilon_y$ are the dielectric permittivities along $x$ and $y$, and $a$ is the distance between the center and the vertex of the hyperbola (Supplementary Information). Fig. 4a shows numerical simulations of the electric field Re$[E_z(x, y)]$ on a α-MoO$_3$ slab using a retroreflector with a shape given by Eq. (1) for HPhPs in α-MoO$_3$ at an incident frequency of $\omega_0 = 920$ cm$^{-1}$ and a distance $a$ equal to $\lambda_p = 1.7$ µm, where $\lambda_p$ is the analytical polariton wavelength along the x direction[34] ([100] crystal direction). All the incident waves reflect back to the location of the source, yielding an enhancement of the intensity at the source of ~+35%. Fig. 4B shows a comparison with a standard reflector, such as a straight edge, in which HPhPs propagate far away from the source (red arrows in Fig. 4b). By mirroring the hyperbolic retroreflector structure with respect to the location of the source, we design an optical nanoresonator for in-plane HPhPs (Fig. 4c) in which all the hyperbolic waves launched by a dipole located at its centre continuously reflect back between its boundaries. In this case, the intensity at the source is enhanced ~+64%. While this value could be further improved by making the cavity dimensions smaller, here we introduce a first proof-of-concept hyperbolic nanoresonator which exploits both the strong directionality of in-plane HPhPs and their high-momenta to enhance light-matter interactions and hence with outstanding prospects for nanophotonics[35].

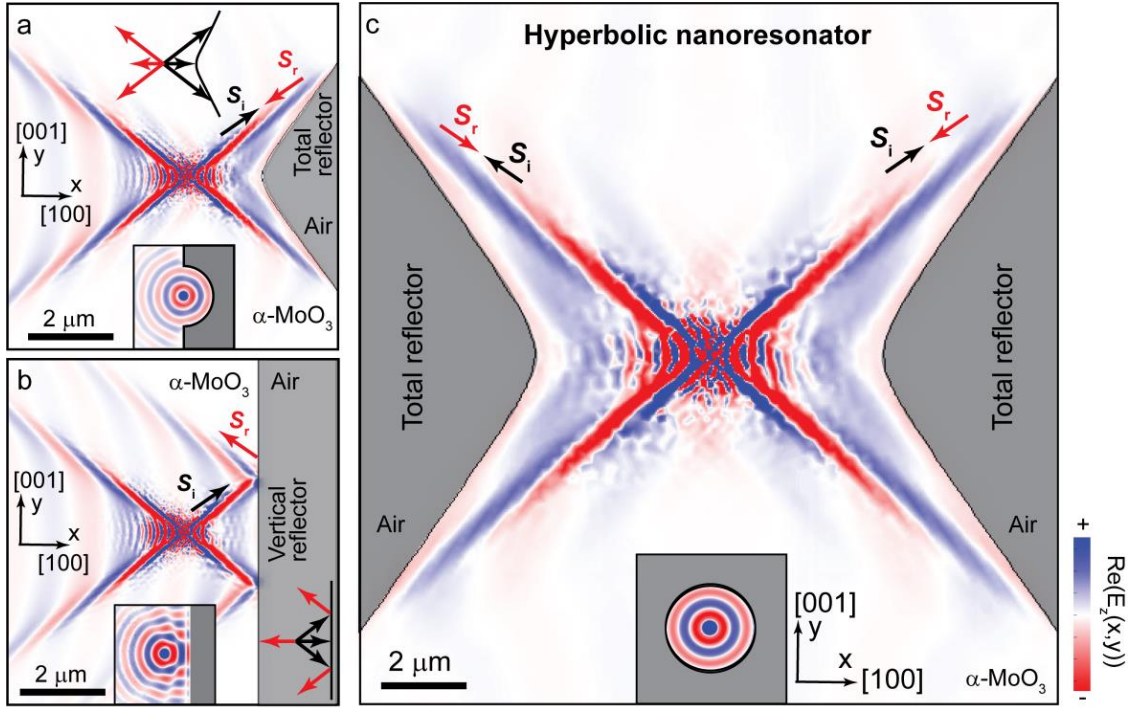

**Fig. 4 | Retroreflector and nanoresonator for HPhPs.** Simulated near-field image $Re[E_z(x,y)]$ of HPhPs reflecting on (**a**) a hyperbolic retroreflector, (**b**) a straight edge, and (**c**) in a hyperbolic nanoresonator consisting of two retroreflectors, on a 200-nm-thick $\alpha$-MoO$_3$ slab at an incident laser frequency of $\omega_0 = 920$ cm$^{-1}$. The source is a point dipole located at a distance equal to the HPhP wavelength $\lambda_p = 1.7$ μm from the edge. The insets show $Re[E_z(x,y)]$ of isotropic polaritons upon reflection on (**a**) a retro-reflector in isotropic media (a semi circumference), (**b**) a straight edge, and (**c**) a nanoresonator in isotropic media (a circumference).

In conclusion, we have demonstrated negative reflection of nanoscale-confined HPhPs in a low-loss natural medium ($\alpha$-MoO$_3$). Our study unveils an unconventional tunability of the direction of propagation and wavelength of the negatively reflected HPhPs as a function of the incident frequency $\omega_0$ and the angle of the mirror $\varphi$, opening a plethora of possibilities to control nano-light by reflection in hyperbolic media. These fundamental insights allowed us to develop a novel optical nanodevice: a hyperbolic nanoresonator. Due to the high density of optical states in hyperbolic media, these cavities hold promises for controlling the decay rates of emitters. Also, open cavities as our hyperbolic nanoresonator are more accessible from the outside, and thus experimentally preferred to design a photonic circuit. Altogether, our findings provide a new route for controlling and manipulating the flow of energy at the nanoscale by employing hyperbolic waves, as well as for squeezing light to nano- or sub-wavelength scales, thereby enabling to compress and enhance light-matter interactions to unprecedentedly small volumes, the general premise of nanophotonics. Our results thus hold promises for the development of applications in optoelectronics, biosensing, nanoimaging and chemistry.

**Author contributions**

**Competing interests:**